\documentclass[aps,prc,groupedaddress,twocolumn,showpacs,floatfix,nofootinbib]{revtex4-1}
\usepackage{graphicx}
\usepackage[dvips]{color}
\usepackage{colordvi}

\def\gtorder{\mathrel{\raise.3ex\hbox{$>$}\mkern-14mu
 \lower0.6ex\hbox{$\sim$}}}
\def\ltorder{\mathrel{\raise.3ex\hbox{$<$}\mkern-14mu
 \lower0.6ex\hbox{$\sim$}}}

\def\beq{\begin{equation}}
\def\eeq{\end{equation}}
\def\ba{\begin{eqnarray*}}
\def\ea{\end{eqnarray*}}

\newcommand{\et}{{\em et al.}}

\begin{document}

\title{Proton rms-radii from low-q power expansions?}

\author{Ingo Sick and Dirk Trautmann}
\affiliation{Dept.~f\"{u}r Physik, Universit\"{a}t Basel,
CH4056 Basel, Switzerland}

\date{\today}
\vspace*{5mm}

\begin{abstract}
Several recent publications  claim that the proton charge {\em rms}-radius 
resulting from the analysis of electron scattering data restricted to {\em low} 
momentum transfer agrees with the radius determined from muonic hydrogen, in contrast to 
the radius resulting from analyses of the full (e,e) data set which is $0.04fm$
 larger. Here we show why these publications erroneously arrive at the low radii. 
\end{abstract}

\pacs{14.20Dh,21.10.Ft,25.30.Bf}

\email{Ingo.Sick@unibas.ch, Dirk.Trautmann@unibas.ch}
\maketitle

 {\em Introduction.~~}  The determination of the  {\em rms}-radius $R$ of the
proton charge distribution has recently attracted much attention. While standard
analyses of electron-proton scattering data yield $0.879 \pm 0.009fm$
\cite{Arrington15}, the Lamb shift measurement in muonic hydrogen gave $0.8409
\pm 0.0004fm$ \cite{Pohl10a}; this represents a $\approx 5\sigma$ discrepancy.  The
radii from electron scattering near $0.88fm$ come from analyses that fit with
excellent $\chi^2$ the {\em world} cross section and polarization transfer data
up to large momentum transfer $q$, $5fm^{-1}$ to $12fm^{-1}$    
\cite{Sick12,Lee15,Borisyuk10,Graczyk14,Bernauer10b,Bernauer14}.   Recently, 
3 publications
\cite{Griffioen16,Higinbotham16,Horbatsch16} which restrict the analysis to the
{\em low}-$q$ data, with $q_{max}=0.7, 0.9$ and $1.6fm^{-1}$ respectively,  
find $R$ in the $0.84fm$ neighborhood, {\em i.e.} compatible with the radius
from muonic hydrogen.  In this paper, we show why these analyses, which yield
values of $R \approx 0.04fm$ lower than refs. 
\cite{Sick12,Lee15,Borisyuk10,Graczyk14,Bernauer10b,Bernauer14}, have led to 
erroneously low  values.

 {\em Power series expansion.~~}   In terms of the electric Sachs form
factor $G_e(q)$ the proton charge  {\em rms}-radius $R$ is defined via the 
slope of $G_e(q^2)$ at $q^2=0$.
It therefore seems natural to parameterize $G(q)$ in a power series 
\beq
G_e(q) = 1 + q^2 a_2 +q^4 a_4 +q^6 a_6 +...
\eeq
where $R^2 = -6a_2$. Non-relativistically, $a_4 = \langle r^4 \rangle/120$ and
$a_6 = -\langle r^6 \rangle /5040 $ are given by the higher moments of the
charge density distribution. 
The rationale behind an analysis restricted to data with {\em low} maximum 
momentum  transfer $q_{max}$: at low
enough $q$ the terms proportional to $q^{2n}$ with $n>1$  (or in some cases
$n>2$) can be neglected, so a linear (quadratic) fit of the data in terms of
powers of $q^2$  should suffice. Low order (one parameter) fits in terms of 
derived functions  as {\em e.g.} a dipole, $G(q) = 1/(1+q^2 b_2)^2$, 
follow the same rationale, although these parameterizations do implicitly
contain higher $q^{2n} a_{2n}$ contributions as fixed by the analytical shape of
the parameterization.

Problems with  expansions of the proton form factors in terms of $q^{2n}$ have
been recognized earlier\cite{Sick03c}. Due to the peculiar shape of the proton
form factor --- approximately a dipole --- and the peculiar shape of the
corresponding charge density --- approximately an exponential --- the moments
$\langle r^{2n} \rangle$ for $n \ge 2$ grow  unusually fast with increasing
order $n$. In the form factor  $G(q)$ the moments $\langle r^{2n} \rangle$ are
tightly coupled and give contributions of alternating signs. In an expansion
with small $n$ ($n=1,2$) the values found for $\langle r^{2n} \rangle$ depend 
on the maximum $n$ and the value of the maximum momentum transfer $q_{max}$
employed, and always yield too small $\langle r^2 \rangle$. This has recently
been shown  by Kraus \et \cite{Kraus14} who quantitatively demonstrate the
pitfalls of fits with low  order power series by analyzing  pseudo-data
generated with known $R$. They  show that {\em e.g.} a linear fit in $q^2$ with
$q_{max} = 0.7fm^{-1}$ as employed in \cite{Griffioen16,Higinbotham16}  
produces a value of $R$ which is low by $0.04fm$.

This result of Kraus \et~can qualitatively be understood. When terminating the
series eq.(1) with the $q^2$-term, one implicitly posits $\langle r^4 \rangle =
0$. As  $\langle r^2 \rangle \approx 0.7fm^2$ this implies a charge density that is
positive at small $r$ (charge proton $+e$), but has a negative tail at large
$r$;  due to the larger weight in the $r^4$-term the tail can reduce $\langle
r^4 \rangle$ to 0. This negative tail of course also affects $\langle r^2
\rangle$, and leads to the systematically low values of $R$. The same happens
{\em mutatis mutandis} with truncations at higher order\cite{Kraus14}.

The second, obvious, problem with very low $q$: the finite size effect (FSE)
$1-G_e(q)$ decreases like $q_{max}^2$. Already at the $q \approx  0.8fm^{-1}$ of
maximal sensitivity of the data to $R$  (see below) the FSE $\approx q^2 R^2 /6$
amounts to 0.09 only. The smallness of the FSE emphasizes that fits used to
extract $R$ must reach the minimal $\chi^2_{min}$ achievable, a visually good 
fit is not enough:  a change of $R$ of 1\% corresponds to
a systematic   change of $G_e$   of only 0.0015 (0.17\% of $G_e$), a difference 
that is  far below  the resolution of typical plots of $G_e (q)$
\cite{Higinbotham16,Griffioen16,Horbatsch16}.

The sensitivity of the data to $R$ is shown in {Fig.1} which results from a  
notch test employing  SOG fits of the {\em world} data   (for recent reference to
notch tests see \cite{Yang16}).     When  exploiting only
part of the  range of  $q \le 1.5fm^{-1}$, one looses part of the  experimental
information on $R$; analyses which limit the data to {\em e.g.} \mbox{0.8
$fm^{-1}$} as done in refs.\cite{Higinbotham16,Griffioen16} then ignore half of
the data sensitive to $R$. Restriction to a subset of the {\em world} data only
amplifies this problem.  

\newpage
\begin{figure}[tbh]
\begin{center}
\includegraphics[scale=0.50,clip]{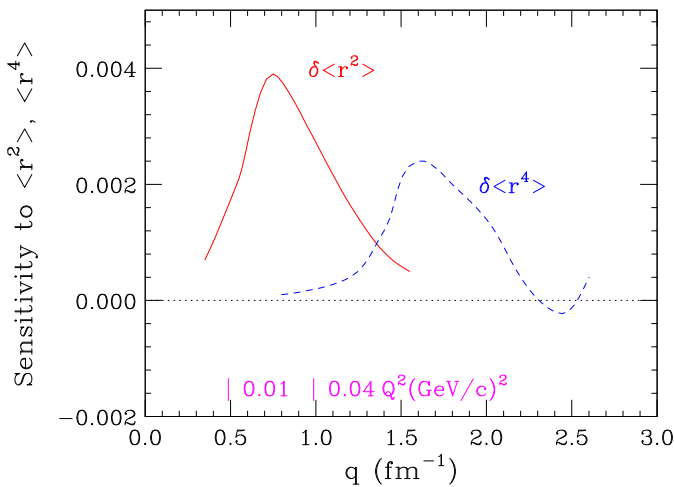}
\parbox{8cm}{\caption[]{(Color online) Sensitivity (arbitrary units) to the 
moments $\langle r^2 \rangle$ and
$\langle r^4 \rangle$ obtained from fits of the {\em world} data.
 }} 
\end{center} 
\end{figure}

{\em Contribution of higher moments.~~} For a more detailed discussion of the
problems with eq.(1), we start from the values of $a_2, a_4,... $ determined by
Bernauer  \et~\cite{Bernauer10a} via a power-series fit (with a $\chi^2$ as 
low as a spline fit) to the Mainz data for
$q_{max} = 5fm^{-1}$.  One might hope that, due to the large $q_{max}$ and
the high order $2n=20$ employed,  the values of the lowest moments  of interest
here should not be affected seriously by the above-mentioned
problems \cite{Sick03c}. Fig.2 shows the percent contribution of the  $a_4$
to $a_{10}$ terms to the FSE. Also indicated is the  uncertainty in the FSE due 
to a (very optimistic) uncertainty of 0.2\% in the  experimental $G_e(q)$.

This figure shows several features: 

1. At the $q$'s  used in the
`low-$q$ fits' referred to above, with $q_{max}=0.72-0.9fm^{-1}$, the 
contribution of the
$q^4$-term to the FSE $\approx q^2R^2/6$ amounts to 10--15\% at the upper limit 
of the $q$-range where FSE is most sensitive to $R$. This shows immediately and
 without further
calculation that neglecting  this contribution in a linear fit in terms of $q^2$
must yield  a value of $R^2$ which is low by a comparable  percentage.  

2. Even the
contribution of the $q^6$-term is not entirely negligible (15\% of the
$q^4$-term at $q = 0.9fm^{-1}$); when attempting to determine $a_4$
from a fit quadratic in $q^2$ a wrong value results if the contribution of the
$q^6$-term  is not accounted for. 

3. Restriction of $q_{max}$ to extremely low
values,  such as to justifiably neglect  the $q^4$-term and 
maintain an
accuracy of 1\% in $R$, would require $q_{max} < 0.35fm^{-1}$. At these values
 of $q$, the FSE
is $<0.015$, and the typical error  bars of $G_e(q)$ would yield huge uncertainties
in the FSE contribution, hence  $R^2$ (see dashed curve).

\begin{figure}[htb]
\begin{center}
\includegraphics[scale=0.50,clip]{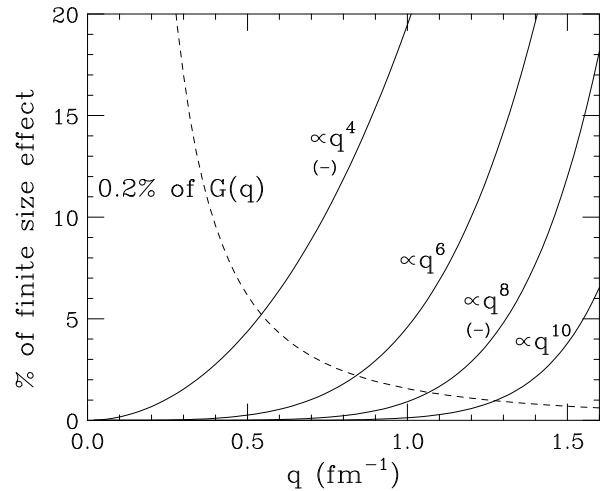}
 \parbox{8cm}{\caption[]{The  solid curves show the relative contribution
(in \%) of the $q^{2n}$ terms  to the finite size effect FSE in 
$G_e(q)$. The dashed
curve shows the relative contribution of an 0.2\% uncertainty of the
experimental $G_e(q)$. For comparison: the $q_{max}$ of the fits linear in
$q^2$ (dipole) of refs. \cite{Griffioen16,Higinbotham16}(\cite{Horbatsch16}) 
amount to  0.72, 0.90 and 1.6$fm^{-1}$, respectively. }} 
\end{center} 
\end{figure}

Fig.2 makes it   obvious that the low-$q$ fits of
refs.\cite{Higinbotham16,Griffioen16}, which neglect the $q^4$-contribution,
must find wrong values for $R$ due to the omitted  $q^4$ term (for a
quantitative analysis see below). 
Fig.2 also shows, without further calculation,  that for $q \leq 1.6fm^{-1}$ 
the information content of the
data is 4-5 parameters (moments) which hardly can be represented correctly by
a one-parameter  formfactor such as employed by 
Horbatsch+Hessels\cite{Horbatsch16} (for a quantitative discussion see below). 

{\em Higher moments from world data.~~} As was pointed out in
\cite{Sick03c} and quantitatively demonstrated in \cite{Kraus14} the
determination of the lowest moments via a power-series fit is not very reliable
and for the higher $n$  dependent on the cut-off in $n$. We
therefore have made an independent determination.

We use the {\em world} data up to the maximum momentum transfer available for
$G_e$, 10$fm^{-1}$ (not including the data of ref.\cite{Bernauer10a} which show
systematic differences \cite{Sick12}). This data set, which comprises 
603 cross sections and polarization
transfer points, is corrected for 2-photon exchange effects \cite{Blunden05}
and fitted with  a Fourier transform of Laguerre functions of order 11 for
both $G_e (q)$ and $G_m (q)$. Laguerre functions\footnote{For similar
expansions see \cite{Kelly02,Friar73a,Anni95}} are particularly well suited 
 as \\
-- They provide an  orthonormal basis which makes multi-parameter fits very
efficient   (even if the polynomials are not strictly orthogonal over the
limited $q$-range of the data).    \\
-- They have a controlled behavior at large radii $r$ due to the $e^{-\gamma r}$
weight function, a consideration which is particularly important \cite{Sick14}
 when addressing 
higher moments (an aspect shared with the parameterizations of the Vector Dominance
Model VDM).\\
-- They provide values for the moments insensitive to the cutoff in the number 
of terms employed; the moments $\langle r^{2n} \rangle$ are given by the lowest
$2n+3$ coefficients. 

The set of data can be reproduced with a $\chi^2$ of 542 with 548  degrees of
freedom when the normalizations of the individual data sets are floated. When
keeping the normalizations at their measured values, and {\em without}
increasing the error bars due to systematic error of the normalizations, the
$\chi^2$ amounts to 783 with 580 degrees of freedom. These $\chi^2$ values are
excellent given a set of data measured over some 50 years. The resulting values
for $\langle r^4 \rangle$ are $2.01\pm0.05~ (1.99)fm^4$.   The quality of the
fit and the values of the  moments are very close to the ones obtained using 
SOG \cite{Sick72b}    ($\langle r^4 \rangle = 2.03)$    or  a VDM-type
parameterization   ($\langle r^4 \rangle= 2.01$). We have verified that a
variation of $q_{max}$ between 7 and 12$fm^{-1}$ and a variation of $n$ between
10 and 13 changes $\langle r^4 \rangle$ by $<0.03 fm^4$. Distler \et
\cite{Distler11a} obtained 2.59$\pm 0.19 \pm 0.04$ from a mix of two form factor
parametrizations fit separately  to low-$q$ \cite{Bernauer10a} and high-$q$
\cite{Arrington07} data.   With these preliminaries we are  in the position to
quantitatively discuss the recent low-$q$ fits.

{\em Fits to very-low $q$ data.~~} Higinbotham \et~\cite{Higinbotham16} perform a linear fit
in $q^2$ to a  subset of the data available, the form factors of
Mainz80+Saskatoon74\cite{Simon80,Murphy74}. For their highest $q_{max}$ of
$0.9fm^{-1}$, which yields the result with the smallest uncertainty, they 
find\footnote{Including Coulomb distortion would have increased $R$ by 
\mbox{$\approx 0.01fm$} \cite{Rosenfelder00}} $R=0.844 \pm 0.014fm$. From this the
authors conclude that $R$ agrees with the value of $  0.84fm$ from muonic
hydrogen.  When repeating exactly the same analysis, but adding in the $q^4$ and
$q^6$ contributions using the   higher moments from the fit to the high-$q$ 
data, one finds a reduced $\chi^2$ ({\em i.e.} $\chi^2$ per degree of freedom)
which is 11\% smaller and a radius $R$ of $0.899fm$. This $R$ disagrees with the
muonic value, and agrees with the above-cited $R$'s in the $0.88fm$ region.

Higinbotham \et~also perform a fit quadratic in $q^2$, and find a radius of
$0.873\pm0.039fm$.  This agrees   with the radii in the $0.88fm$ region,
although, as the authors want to see it, the value is ``within one $\sigma$ of
the muonic result''. The uncertainty of $\pm0.039fm$  illustrates  the large
error bars resulting from the restriction of the analysis to a fraction of the
$q$-region sensitive to $R$ (see Fig.1)  and the large uncertainty of 
$\langle r^4 \rangle$ due to the truncation in $q$.  When
using, instead of the  $\langle r^4 \rangle = 1.32\pm0.96$ of Higinbotham \et ,
the value $2.01 \pm 0.05$ we know from the fit to the high-$q$ data, the result 
for $R$ becomes $0.901fm$, with a smaller error bar of $0.010fm$.

Griffioen \et~\cite{Griffioen16} analyze part of the cross sections of 
\cite{Bernauer10b} for  $q <0.72fm^{-1}$ using eq.(1) including terms up to
$a_4$. They use a  low-$q$ parameterization for $G_m/G_e$ and take the shortcut
of ignoring the free relative normalizations of the individual data
sets\footnote{Correct treatment  of the  normalizations of the data 
sets of \cite{Bernauer10b}, which are   individually   floating,  would have 
increased the
uncertainty of $R$ by a factor 1.6.}. They find an $rms$-radius of
$0.850\pm0.019fm$ and conclude that this value is consistent with the muonic
hydrogen result of $0.84fm$. 
Repeating their fit, but using  the  $a_4$ determined much better
from the high-$q$  fit, yields a radius of $0.877\pm0.008fm$, with lower
$\chi^2$  and a significantly smaller error bar.  This result  agrees with the 
$0.88fm$-type results, and disagrees with  the radius from muonic hydrogen.

Griffioen \et~ also perform fits up to order $q^6$, with $a_4, a_6$-values as
given by simple models for the proton charge density (uniform, exponential,
gaussian) which all produce the same $\chi^2$; the resulting R-values are
linearly correlated with $a_4$.   Extrapolating these values linearly to the
value of $a_4$  given by the  fit to high-$q$ data yields \mbox{$R =
0.876\pm0.008fm$}, again in agreement with the $R$'s in the $0.88fm$ region.

The bottom line: all the low-$q$ fits of refs.\cite{Higinbotham16,Griffioen16}
yield radii in the $0.88fm$ region once the higher moments of the charge density
--- which {\em are} non-zero but  ignored (or poorly fixed in the low-$q$ fits
 due to the   
truncation of the series in $n$ of $q_{max}$)   ---  are properly accounted for. 

{\em Fits to not-so-low $q$ data.~~} Horbatsch and Hessels \cite{Horbatsch16} employ the
 cross sections of ref.\cite{Bernauer10b} up to a  $q_{max}$ of $1.6fm^{-1}$.
They parameterize the form factors via a 1-parameter dipole expression for both
$G_e$ and $G_m$. Their fit
yields a reduced $\chi^2$ of 1.11, and a (charge) $rms$-radius 
\mbox{$R = 0.842\pm0.002fm$}. From this, together with other fits which yield 
radii near $0.89fm$, the authors conclude that $R$ is in the range 
$0.84-0.89fm$, {\em i.e.} could be compatible with the radius from muonic
hydrogen. 

 Fig.2 shows that for $q_{max}=1.6fm^{-1}$ the moments up to  at least $2n=10$
are important to get the full FSE. It is highly unlikely that the one-parameter
dipole  contains the mix of $q^{2n}$-terms  for $2n=4...10$
appropriate for the proton. Indeed, expansion of the dipole  in terms
of powers of $q^2$ shows that the numerically largest difference to the 
power-series fit of \cite{Bernauer10a}
results from the  contribution of  the $\langle r^4 \rangle$ term.  This
difference in $\langle r^4 \rangle$ alone  would lead, at the $q=0.85fm^{-1}$
of maximal sensitivity to $R$, to a difference $\Delta G_e$ of
0.0081 corresponding to 9.5\% in the FSE, hence $R^2$ (causing the systematic 
deviations   just   visible in Fig.3 of \cite{Horbatsch16}). The same consideration 
applies to the parameterization of $G(q)$ as a
(one-parameter) linear function $1 - c z$ with 
$z=(\sqrt{t_c-t}-\sqrt{t_c})/(\sqrt{t_c-t}+\sqrt{t_c})$ and $t=-q^2$. The
lacking flexibility of the fit function, causing  systematic differences between
data and fit and a $\chi^2$ larger than the one of already published fits, also 
affects
 the  results  from the  high-$q$ fits of \cite{Higinbotham16,Griffioen16}.

For the fits of Horbatsch and Hessels it is not practical to correct for the
effect upon $R$ of the incorrect higher $q^{2n}$-terms as we did above
for the analyses of refs.\cite{Higinbotham16,Griffioen16}; too many terms
$2n=4...10$ would contribute. In order to demonstrate the importance of their
effect we  rather quote the result of a Laguerre-function fit (4 terms each for
$G_e$ and $G_m$) to exactly the same data, yielding a lower reduced $\chi^2$ of
1.045 and a (charge)  $rms$-radius  $R=0.884\pm0.016fm$. Due to the lacking
flexibility the parameterization  of 
Horbatsch+Hessels has a $\chi^2$ that is higher by 50! From such a  ``fit'',
that is some 7 $\sigma$'s away from a genuine best-fit, one obviously cannot 
get  a significant value for $R$.  

{\em Conclusion.~~} The moments $\langle r^{2n} \rangle$ of the
proton  for \mbox{$n>1$} {\em are} there, and they are {\em known}  to be 
large.  Ignoring 
their strong correlation with $R$ \cite{Higinbotham16,Griffioen16,Horbatsch16}
leads to wrong results for the proton $rms$-radius.

\end{document}